\documentstyle[preprint,aps]{revtex}

\begin{document}
\newfont{\bftitle}{cmbx12 scaled\magstep5}
\newcommand{\be}{\begin{equation}}
\newcommand{\bea}{\begin{eqnarray}}
\newcommand{\eea}{\end{eqnarray}}
\newcommand{\ee}{\end{equation}}
\newcommand{\bb}{\langle}
\newcommand{\kk}{\rangle}
\newcommand{\bk}[4]{\bb #1\,#2 \!\mid\! #3\,#4 \kk}
\newcommand{\kb}[4]{\mid\!#1\,#2 \!\mid}
\newcommand{\kx}[2]{\mid\! #1\,#2 \kk}

\draft

\title{The Role of Color Neutrality in Nuclear Physics-
Modifications of Nucleonic Wave Functions}
\author{M. R. Frank}
\address{
Institute for Nuclear Theory, University of Washington, Seattle, WA 98195, USA}
\author{B. K. Jennings}
\address{
TRIUMF, Vancouver, BC V6T 2A3, Canada}
\author{G. A. Miller}
\address{Dep't of Physics, Box 351560
University of Washington, Seattle, WA 98195-1560, USA}
\maketitle

\today

\begin{abstract}
The influence of the nuclear medium upon the internal
structure of a composite nucleon is examined.  The interaction with the
medium is assumed to depend on the relative distances
between the quarks in the nucleon consistent with the
notion of color neutrality, and to be proportional to the
nucleon density.  In the resulting description the nucleon in matter is a
superposition of
the ground state (free nucleon) and radial excitations.  The effects of the
nuclear medium on
the electromagnetic
and weak nucleon form factors, and the nucleon structure function
 are computed using a light-front
constituent quark model.
Further experimental consequences are examined by considering the
electromagnetic nuclear
response functions.  The effects of color neutrality
supply small but significant corrections to predictions of observables.
\end{abstract}

\section{Introduction}
Nucleons are composite color singlet systems made of quarks and gluons.
Their wave functions consist of  many configurations as illustrated
in Fig.\ref{fig1}.  Some configurations are simple with
only three current quarks, most are more complex with many partons.
Configurations in which three quarks are close together have
been dubbed point-like configurations (PLC).\cite{FS85}  According to
perturbative quantum chromodynamics (pQCD), such configurations
are responsible for
high momentum transfer elastic scattering
reactions\cite{muellerpr,BL80}.
 Furthermore, the effects of gluon emission from closely separated
color-singlet systems of quarks and gluons tend to
cancel.
This means that for processes in which one adds amplitudes before
squaring, i.e. coherent processes, point like configurations
do not interact with surrounding media.

That color neutrality can suppress
interactions is
similar to the cancellation of
interactions which would occur if an electron and positron
would move together at the same position through a charged medium
(see for example Ref.\cite{MM95}).
The color neutrality feature of QCD has been verified. It is
responsible for the scaling of structure functions at low, but not
too low, values of the Bjorken scaling variable
$x_{Bj}$. (See the reviews \cite{FS88,nnn,FMS94}.)
Furthermore hadron-proton total cross sections $\sigma_{hp}$
grow linearly with the mean square radius of the hadron\cite{ph}.

There are also configurations in the nucleon wave function in which the
partons occupy a larger than average size. We call these
configurations blob-like configurations BLC or
huskyons\cite{BF93}.  These configurations have complicated
strong interactions with the medium.  (See \cite{BF93}
which explores the consequences of huskyons.)  Here
we model the interaction of the nucleon with
the surrounding medium in terms of the inter-quark separation within the
nucleon to investigate the
role of color neutrality and its consequences in
nuclear wave functions.

Sixty years of studies of
nuclear properties have  shown that
the nuclear wave function is dominated by clusters of color
singlet objects with the quantum numbers of
nucleons and mesons.  The success of the nuclear shell model
is a testimony to this fact.
It is certainly possible, however, that although bound nucleons
largely maintain their identity, bound and free nucleons
are not identical. Indeed, there are a number of
interesting experimental findings which may indicate that medium
modifications of nucleon properties are relevant.

One of the most spectacular examples is the
observation of the first EMC effect
\cite{EMCrefs} showing that the structure function of the bound nucleon
was suppressed at large $x_{Bj}$.
There have also been  
numerous
studies of the ($e,e'$) reaction which find that the longitudinal
response function $R_L$ is suppressed, while the transverse
response $R_T$ is not. 
 See
Ref.\cite{Zghiche} and references therein.
The suppression of $R_L$ is natural in some theories\cite{ths}.
  This lore has been challenged recently by
Jourdan\cite{jo95} who argues that the ``so-called quenching is
mostly due to the limited significance of the data" and that including
data at high energy loss $\omega$ leads to the result that ``no A-dependent
quenching  is observed". However, the various errors listed
in Jourdan's Table 2 allow for up to about 15\% effects.

It has also long been stated that the nuclear
value of the axial coupling constant $G_A$ was less than its value
in free space\cite{garefs}. Furthermore the large pionic enhancement
expected in many models of nuclear structure was not
observed\cite{pirefs}, but there may be hints of a pionic
enhancement\cite{newprl}.

Thus there are several indications that nucleons could be modified by
the medium. Since color is at the heart of QCD, one might suspect
that looking at the consequences that color neutrality might
have for bound composite nucleons might be worthwhile.  These are the
presumably small effects of nucleonic polarization.  Nevertheless, as we
show here, these effects can have nontrivial consequences.
Before describing in detail our treatment of the nucleon in the medium, it is
necessary to mention some of the possible objections and also to
discuss some of the relevant history.

The most significant objection is that there could be many
effects other than color neutrality which cause
significant modifications of the properties of nucleons. This is
certainly a valid point.  For example,
in the Walecka model\cite{walecka} the medium provides
a decrease of the nucleon mass by as much
as 30\%.  See also the quark-meson coupling model of Ref.\cite{QMC},
and the works of Ref.\cite{mkb,tg}. 
It is entirely conceivable that similar effects to the
nucleon form factors could occur from a purely hadronic mechanism, i.e.,
independent of the relative quark separation.
It has also been argued that a more general scaling of hadronic masses
may arise in the nuclear medium.\cite{brown}  The extension
of QCD sum rules to finite density further confirms the
presence of large scalar and vector self-energy corrections\cite{QCDSR,hp}.  
Nevertheless, these effects have a different origin than that of
color neutrality, and our aim here is to
simply study the consequences of this single effect in several
different situations by identifying
a pattern of predictions. The study of many reactions is necessary, so
we derive a formalism
applicable to many reactions. This formalism may be useful
in the study of other mechanisms for medium modifications as well as
other corrections to the impulse approximation.

Another objection is that
it is not clear when pQCD, and hence its
discussion of point-like configurations
is applicable.
The question of how high
the momentum transfer must be for pQCD considerations to dominate
calculations of form factors is controversial. The work of
Refs. \cite{ar,isgurls} is merely the first example of a long history.
More data are needed to settle this question. However, point-like
configurations
may arise from
non-perturbative considerations as well\cite{FMS92,FMS931}. The singular
nature of the non-perturbative confining interaction can lead to
point-like configurations.
Indeed, many constituent quark and Skyrmion models seem
to indicate the presence of point-like configurations.
It is therefore of interest
to consider an interaction with the nuclear medium which
accomodates such objects.

Another possible objection is the failure to observe clear and
convincing evidence for color transparency CT. This is the idea that
a high $Q^2$ quasielastic reaction produces a
point-like configuration
which does not interact with the surrounding nuclear medium.
In this way initial and/or final state interactions are reduced.
 A BNL experiment\cite{bnl88}
shows some evidence for CT in a ($p,pp$) experiment at beam
momenta of 6, 10 and
12 GeV/c (4.5 GeV$^2<Q^2<8$ GeV$^2$) but the
$(e,e'p)$ NE18 ($Q^2=1,3,5,7$ GeV$^2$)
experiment at SLAC\cite{ne18} sees no such evidence.
A recent FNL exclusive $\rho $ production experiment saw evidence
for color transparency\cite{FNL}. The Q$^2$ was up to about 9 GeV$^2$.
The BNL and FNL experiments may not have guaranteed the necessary
quasielastic nature of the experiment. The SLAC  experiment did
that.  One interpretation of these results is that
point-like configurations
are produced but expand before leaving the nucleus.
This expansion is a bigger effect for the NE18 experiment, which has the lowest
energy outgoing protons. Expansion is also a big effect for the BNL experiment,
but the
incident proton has high enough momentum for this expansion not to completely
kill the influence of the point-like configurations.
 The  expansion effects are smallest for
the  FNL experiment, but the experimental resolution may be a problem.

None of the considerations of the above paragraphs are sufficient
to rule out the existence of point-like configurations,
 or the effects of color neutrality
in nuclear physics. Indeed,  
point-like configurations were introduced in the context of the EMC
effect\cite{FS85}.  Frankfurt and Strikman postulated that such configurations
do not feel the attractive nuclear interaction and therefore appear in a
smaller percentage in nuclei than in free space. This leads to a
reduction of the nuclear structure function at $x_{Bj}\sim
0.4-0.6$. This effect also reduces the cross sections predicted for ($e,e'p$)
reactions,
making color transparency harder to observe \cite{FSZ93}.
Frankfurt and Strikman
treat their effect by taking the suppression to
vary essentially as a theta function in $Q^2$, i.e., present at
$Q^2$ above a certain value but completely absent for lower
values.  Desplanques finds a similar term by including the  spatial variation
of the nuclear scalar and vector mean-fields
 over the volume of the nucleon\cite{desplanques}.
Desplanques's treatment was non-relativistic.
Note also that 
Jain and Ralston\cite{ralston} have argued
that hard processes are modified significantly  by medium effects.

Our aim here is to present a method for doing calculations which
may be applied both at high and low values of Q$^2$. We also
obtain a Lorentz covariant formulation. Here is an outline.
Our model and basic formalism
are presented in  the next section. Section 3 deals with some
simple examples which allow us to study our approximations and expose
the need for a relativistic
formalism. The relativistic formalism, which applies the
recent work of Schlumpf\cite {fs1,fs2,fs3},  is discussed next. Section 5 is
concerned with a presentation of our detailed numerical results.
We study medium modifications of the electromagnetic proton form
factors $G_E$ and $G_M$ and the axial vector form factor $G_A$
relevant in the weak interaction. We discuss consequences
in the $e,e'$ and $e,e'p$ reactions.
A concluding section discusses and asseses these results.

\section{Formalism and Model}

We begin from general considerations. Suppose a color singlet baryon moves
in
the nucleus subject to a Hamiltonian
$H$ given by
\begin{equation}
H=H_0+ H_1\label{f1}
\end{equation}
where $H_1 $ is a perturbing Hamiltonian, and
\begin{equation}
H_0=H^{cm}_0+H^{rel}_0.\label{f2}
\end{equation}
Here $H^{cm}_0$ and $H^{rel}_0$ describe
respectively the motion of the center of mass
(including effects
of the medium on the center-of-mass motion), and the relative motion of
the internal degrees of freedom independent of the medium and the
center-of-mass motion. The perturbation $H_1$ describes mixing between
the center-of-mass and
relative motion, and thereby incorporates the effects of the medium on the
internal wave function.

We shall assume that $H_1$ is separable, that is,
\begin{equation}
H_1=H^{cm}_1h^{rel}_1 , \label{f3}
\end{equation}
where $H^{cm}_1$ and $h^{rel}_1$ act only
 in the center-of-mass and relative motion
spaces respectively.  Our separation of the Hamiltonian in
(\ref{f1}) and (\ref{f2})
entails that $h_1^{rel}$ requires
a change in the internal structure of the baryon.  This implies that
$h_1^{rel}$ has no diagonal elements, and that the baryon in the medium
is described as a superposition of excitations.

We wish to consider the
 eigenstates $|\Psi_{nm}>$ of the hamiltonian $H$ in (\ref{f1}),
defined by
\begin{equation}
H|\Psi_{nm}> = E_{nm}|\Psi_{nm}>. \label{f4}
\end{equation}
The form of $H_0$ in
(\ref{f2}) allows the unperturbed states to be written as a
direct product of center-of-mass and relative motion states as
\begin{equation}
|\Psi^{(0)}_{nm}>=|\Phi _n>|\phi _m>. \label{f5}
\end{equation}
Here $|\Psi^{(0)}_{nm}>$ satisfies the eigenvalue equation
\begin{equation}
H_0|\Psi ^{(0)}_{nm}> = E^{(0)}_{nm}|\Psi ^{(0)}_{nm}>, \label{f6}
\end{equation}
where $H^{cm}_0|\Phi _n>=\xi_n|\Phi _n> $ and
$H^{rel}_0|\phi _m>=\epsilon_m|\phi _m> $, with
$E^{(0)}_{nm}=\xi_n+\epsilon_m$.

The
state vector of the full Hamiltonian can
be written to first order in $H_1$ as
\begin{equation}
|\Psi _{nm}> = |\Psi ^{(0)}_{nm}>+\sum _{kl\neq nm} \frac{<\Psi ^{(0)}_{kl}|
 H_1|\Psi ^{(0)}_{nm}>}{E^{(0)}_{nm}-E^{(0)}_{kl}}|\Psi ^{(0)}_{kl}>
.\label{f12b}
\end{equation}
To simplify our notation
we restrict our discussion to the nuclear
ground state.  For this case Eq.(\ref{f12b}) can be rewritten as
\begin{equation}
|\Psi _{00}> =|\Psi ^{(0)}_{00}>+\sum_k |\Phi _k><\Phi _k|H^{cm}_1|\Phi _0>
\sum _{l\neq 0}\frac{ <\phi _l| h^{rel}_1 |\phi _0>}{E^{(0)}_{00}-E^{(0)}_{kl}}
|\phi _l>. \label{f14}
\end{equation}
It should be noted that in (\ref{f14}), since $h^{rel}_1$ has no diagonal
elements, $l\neq 0$ and the sum on $k$ is
unrestricted.
The energy denominator in Eq.(\ref{f14}) is dominated by
nucleonic excitation energies. These, which are typically hundreds
of MeV, are much larger than the nuclear energy differences which
are typically tens of MeV. Thus we write
\begin{equation}
E^{(0)}_{00}-E^{(0)}_{kl}\approx\epsilon_0-\epsilon_l.
\end{equation}
This allows the sum on $k$
to be performed using completeness so that
\begin{equation}
|\Psi _{00}> = \left[ |\phi_0>+
\sum _{l\neq 0}\frac{ <\phi _l| h^{rel}_1 |\phi _0>}{\epsilon_0-\epsilon_l}
|\phi _l>H^{cm}_1\right] |\Phi_0>. \label{f15}
\end{equation}
We define the quantity in brackets as $|\tilde{\phi}>$
\begin{equation}
|\tilde{\phi}>\equiv |\phi_0>+\sum _{l\neq 0}
\frac{ <\phi _l| h^{rel}_1 |\phi _0>}
{\epsilon_0-\epsilon_l}|\phi _l>H^{cm}_1, \label{fnew2}
\end{equation}
which
can be regarded as the modified nucleonic wave function. Note that this wave
function depends on the coordinates of the entire nucleus through the operator
$H^{cm}_1$. 
The use of Eq.(\ref{fnew2}) allows one to evaluate the effects of color
neutrality for finite nuclei; this equation
is our principle formal result.

The evaluation of Eq.(\ref{fnew2}) depends on knowing the wave functions
corresponding to the Hamiltonian $H^{rel}_0$. At the present time there
is no complete relativistic treatment available, but
progress has recently been made in this direction\cite{Simula}. 
We wish to obtain an
alternate method of evaluating the influence of the nuclear
medium on nucleonic wave functions. Such can be obtained using a
closure approximation in which all of the strength is assumed (on
average) to lie at an average excitation energy. We use
 $\epsilon_0-\epsilon_l=\Delta E.$
The expectation value
of an operator ${\cal O}$ can then be written
to first order in $H_1$ as
\begin{equation}
<\tilde{\phi} |{\cal O}|\tilde{\phi} >=
<\phi _0|{\cal O}|\phi _0> +
{H^{cm}_1\over\Delta E}
<\phi _0|\left\{  h^{rel}_1,{\cal O}\right\}|\phi _0>,\label{f21}
\end{equation}
where $\{ A,B \} =AB+BA$.
The right side of (\ref{f21}) is independent of the excited state wavefunction,
and can therefore be evaluated based on knowledge of the ground state
wavefunction and the excitation energy.  Thus the uncertainty in
the knowledge of the wave functions is replaced by the
uncertainty in a single parameter, $\Delta E$, and by the further
assumption that $\Delta E$ does not depend very strongly on the
operator ${\cal O}$, i.e. that $\Delta E$ is essentially process
independent. The accuracy of this   approximation and assumption
is investigated in section 4.
We shall discuss reasonable values of $\Delta E$
after discussing our choice of the operator $  h^{rel}_1$.

Equation (\ref{f21}) has a simple physical interpretation.  Consider,
for example, the case where the operator ${\cal O}$ is the electromagnetic(EM)
current $j_{\mu }$ associated with a constituent of the state $|\phi _0>$.
A modification of the form factor will occur by the mechanism illustrated in
Fig.\ref{fig2}.  The current acting on the ground state excites an internal
mode.  The excitation decays back to the ground state via the interaction with
the medium provided by $h^{rel}_1$.  The extent to which the form factor is
modified is determined by the ability of the EM current to produce an
excitation and the ability of the medium to absorb its decay
and return the composite particle to its ground state.  The modification
is directly accountable to the density of the medium in the vicinity
of the interaction through the potential,
$H^{cm}_1(R)$, and is suppressed by the
excitation energy $\Delta E$.

Medium effects
are often estimated by taking the nucleus to be infinite nuclear matter, or
by using a local density approximation.
Eq.(\ref{f21}) shows that instead one can evaluate the
appropriate nuclear
matrix which depends on $H^{cm}_1$.

\section{Color screening}

Consider, now the motion of a composite color singlet baryon
through a nucleus. We are here interested in the properties of
the ground and low lying nuclear states, so that the baryon is
part of a bound  state wave function.
The configurations of
the baryon are pictured in Fig.\ref{fig1}. Let the displacement of the center
of
mass of the baryon from the nuclear center be denoted as $\vec R$.
The interactions between  such a complicated system and the
remainder of the nucleus must depend on the
positions $\vec r_i$ of the partons inside  the baryon.

In general the interactions between a nucleon and the
rest of the nucleus are complicated. Here we are studying
the presumably small effects of nucleonic polarization, so we
concentrate on the necessary modifications of the central part of
the nuclear shell model potential. This is the largest
interaction to consider, and a small modification of it might have
significant consequences. Thus we have the central potential
$V=V(\vec r_i,\vec R)$. What else do we know? Color neutrality
tells us that $V=0$ when $\vec r_i=\vec R$ for all partons i.
Furthermore interactions vanish as $r^2$, where
$r^2=\Sigma_{i<j}(\vec r_i-\vec r_j)^2$\cite{Low75,gs77,Nus75}. Finally we
note that the nucleonic average over $\vec r_i$ should correspond
to the standard nuclear shell model potential. These
considerations allow us to write
\begin{equation}
V(r,R)=V_0\rho(R){r^2\over<r^2> }, \label {assum}
\end{equation}
where $<r^2>\equiv<\phi_0|r^2|\phi_0>$
 is the nucleonic expectation value of the operator
$r^2$, $V_0\approx-50 $MeV and $\rho(R)$ is the nuclear density
normalized so that $\rho(R=0)=1$. 

We are concerned with fluctuations, so that it is convenient
to rewrite the central shell model potential as
\begin{equation}
V(r,R)=V_0\rho(R) +
V_0\rho(R)\quad{r^2-<r^2>\over<r^2> },\label{V(r,R)}
\end{equation}
so that the operator $H_1$ of previous sections
can be identified as
\begin{equation}
H_1=V_0\rho(R)\quad{r^2-<r^2>\over<r^2> }.
\label{H1}
\end{equation}
This is the simplest form of $H_1$ that we can write, which is
consistent with known properties. The behavior for large values
of $r^2$ is simply a guess, but it is reasonable to expect that
large, blob-like configurations should have strong interactions
with the medium.

The equation (\ref{assum}) includes the effects of color neutrality on the
total nuclear mean field and makes no distinction between the
scalar and vector mean fields. The
effects of color neutrality on the
separate fields can be examined using the quark-meson coupling model
\cite{QMC}, but this would require the extension of the applicability of that
model to high values of the transferred momenta.

We note that $H_1$ has the interesting property that
\begin{equation}
<\phi _0|H_1 |\phi _0>=0, \label{fnew1}
\end{equation}
which means that the color neutrality effect
governed by our perturbing Hamiltonian does
not give a first-order shift in the mass of the baryon.  Such
effects are by definition contained in the first term of Eq.(\ref{V(r,R)}).

Let us discuss some of the implications
before presenting the evaluation of specific models.
Firstly, consider the value of $\Delta E$. We see that the
operator $r^2-<r^2>$ excites the breathing mode of the
nucleon. Thus it is reasonable to associate $\Delta E$ with the
the energy at which the first resonance of the nucleon occurs.
This is the Roper resonance with \mbox{$\Delta E$=-500 MeV}.
We are concerned with how certain matrix elements  are influenced
by the presence of the nuclear medium, so it is convenient to
define $\delta<{\cal O}>\equiv<\tilde{\phi} |{\cal O}|\tilde{\phi}>
-<\phi _0|{\cal O}|\phi _0>$. Then  using Eq.(\ref{H1}) in Eq.
(\ref{f21}) leads to
\begin{equation}
\delta<{\cal O}>=2\quad{V_0\rho(R)\over\Delta E}
<\phi _0|{\cal O}\quad {r^2-<r^2>\over<r^2> }|\phi _0>.\label{deltao}
\end{equation}
This shows that the nucleon's properties as measured by
$\delta<{\cal O}>$ depend on the position of the nucleon.
The factor $2V_0\rho(0)/\Delta E$ is about +0.2, which shows that
the effects we study here are small but not insignificant.

It is useful to consider examples. First suppose ${\cal O}=r^2$.
Then
\begin{equation}
\delta<r^2>=2\quad{V_0\rho(R)\over\Delta E}
<\phi _0|\quad {r^4-<r^2>^2\over<r^2> }|\phi _0>.
\end{equation}
Writing $r^4=r^2\sum_n |\phi_n><\phi_n|r^2$, 
subtracting the ground
state contribution, and examining the remainder
and using completeness shows that
$\delta<r^2>\quad>0$. Color neutrality leads to an increase in
 the mean square radius of the
nucleon. Note that the same argument can not be used to show that
$\delta<r^4>\quad>0$; indeed this quantity could be negative.
Color neutrality does not correspond to a general scaling of the
nucleon wave function. Each matrix element must be worked out
independently.

Another example to consider occurs in the non-relativistic quark
model if
 ${\cal O}=\mu\equiv\sum_{i=1,3}\mu_3(i)$. In this case the expectation
value of ${\cal O}$ for a spin up nucleon is the magnetic moment of
free nucleon. In the non-relativistic quark model the wave
function is a product of space and
spin-isospin functions. In this case, the integral over the
spatial coordinates vanishes and $\delta <\mu>=0$. Similarly, in
the non-relativistic quark model   the modification to the
electric and magnetic form factors must be the same. It is
necessary to extend these considerations to relativistic models.

It is very interesting to consider the case when ${\cal O}={\cal O}_0
=\delta[(\vec{r}_1-\vec{r}_2)/\sqrt{2}]\delta[(\vec{r}_1+\vec{r}_2
-2\vec{r}_3)/\sqrt{6}]$ for which all of the quarks are at the same
position.  The expectation value of ${\cal O}_0$ is the square of the
wave function at the origin.  In this case the assumption that
$V(r=0,R)$ vanishes is sufficient to give a result, i.e., no additional
assumption about the dependence on $r^2$ for large $r^2$ is needed.
One finds immediately (using Eq.(\ref{deltao}) for example) that
\begin{equation}
\delta <{\cal O}_0>=-\frac{2V_0\rho(R)}{\Delta E}
\left| \phi _0(0)\right| ^2,
\end{equation}
which corresponds to about a 20\% reduction of the square of the wave function
at the origin.  This result is in good agreement with a recent QCD
sum rule calculation by Jin et al.\cite{Jin}.   

It is worthwhile to compare our approach with that of Frankfurt
and Strikman\cite{FS85}. Those authors consider the nucleon $|N>$ as a sum
of various configurations, $|N>=|PLC>+\cdots$. At high $Q^2$ the
form factors are assumed to be dominated by the $|PLC>$ component.
In the medium $|PLC>$ is replaced by $( 1+{H_1\over \Delta
E})|PLC>$. However, $r^2|PLC>=0$, so that the PLC acquires
a factor  of $(1-{V_0\rho(R)\over \Delta E})$. This leads to
$10-20\%$ reductions in form factors, if $Q^2$ is greater than
some value large enough for $|PLC>$ to dominate.  This result is recovered in
our approach if we assume that the only dependence of the form factors
on the (modification of the) nucleon wave function is through
the wave function at the origin.
Even more striking is the
feature that the $|PLC>$ is suppressed, so that one expects
valence
quarks  to carry less momentum in the nucleus. This leads to an
explanation of the EMC effect. Our approach is motivated by the
original work of Frankfurt and Strikman. However, we do not make
the ``all or nothing " assumption in which $|PLC>$ suppression
turns on abruptly. Thus we are concerned with making more detailed
evaluations than and testing the assumptions of  the early work.
In particular, we wish to learn if these suppression effects persist
if one uses more detailed models.

\section{Toy models}

To gain some insight into the results obtained for the nucleon
in the relativistic framework to follow in section 5,
it is useful to apply the formalism of the
previous two sections to some simple non-relativistic constituent quark
models involving (mainly) only two quarks.  These simplifications
enable us to obtain exact solutions and therefore to assess the
validity of our approximation scheme.
We shall start by using the harmonic oscillator model. However, such a model
is not expected to be a reasonable guess for large momentum
transfers. Indeed previous work \cite{FMS92,FMS931} showed that the
harmonic oscillator does not have a point-like configuration.
Therefore we also consider the case of two quarks bound by a Coulomb
potential. This model assumes that the attractive nature of the
color electric force dominates over the longer range confining
force. Finally we consider a model with both a harmonic
confinement and Coulomb term as in \cite{FMS92,FMS931}. 

\subsection{Harmonic Oscillator}
Here we  start by
considering  the ``nucleon" to be made of two quarks,
 and employ the nonrelativistic
constituent quark model with $H^{rel}_0$ given by the harmonic oscillator form
\begin{equation}
H^{rel}_0=\frac{p^2}{2\mu }+\frac{1}{2}\mu \omega_0 ^2r^2. \label{t1}
\end{equation}
We study the nuclear medium-modified nucleon wave function $\tilde\phi$ at
$R=0$. This
is  given by
\begin{equation}
\left[ \frac{p^2}{2\mu }+\frac{1}{2}\mu
\omega_0^2r^2 +V_0\rho(0){r^2\over <r^2>}\right]
\tilde\phi=E\tilde\phi.
\end{equation}
We may easily obtain the exact solution by realizing that in the medium the
frequency $\omega_0$ is replaced by $\omega$ with
\begin{eqnarray}
\omega^2=\omega_0^2+{2 V_0\rho(0)\over \mu <r^2>} \nonumber\\
=\omega_0^2+{4\over 3} V_0\rho(0)\omega_0.
\label{omega}
\end{eqnarray}
For
typical values of $\Delta E=-2\omega_0\sim -500$ MeV and
$V_0=-50$ MeV, we find that $\omega=214$ MeV and the expectation value of $r^2$
increases by a ratio of 
1.17. So, at the nuclear center,
the effect of color
neutrality can be significant. The expectation value of $r^{-1}$ varies as
the inverse of the square root of the frequency, so that
its expectation value is decreased by a factor of 0.93.

We may also compute the form factor for this model.
This is the
Fourier transform of the square of the wave function as a
function of the momentum transfer
$Q=|\vec Q|$ divided by two:
\begin{eqnarray}
F_0(Q^2)=<\phi_0|e^{i\vec Q\cdot {\vec r\over 2}}|\phi_0>\\
F(Q^2)=<\tilde\phi|e^{i\vec Q\cdot {\vec r\over 2}}|\tilde\phi>.
\end{eqnarray}
 The free form factor $F_0(Q)=e^{-{Q^2\over 16
\mu\omega_0}}$, while the medium modified form factor
$F(Q)=e^{-{Q^2\over 16 \mu\omega}}$. We see that
\begin{equation}
\lim_{Q^2\to\infty} {F(Q^2)\over F_0(Q^2)}=0,
\end{equation}
so that huge effects at high $Q^2$ are possible. However, this
model is not realistic at high $Q^2$.

We may also consider the effects of color neutrality for very dense
nuclear systems
by increasing $\rho(0)$ above its value of unity in normal
nuclear matter.
Eq.(\ref{omega}) tells us  that $\omega$ vanishes for densities
about four times nuclear matter density for which
$\rho(0)={3\omega_0\over 4V_0}=3.75$. Thus there is a ``deconfinement"
phase transition inherent in our model. However, the model is
built on the assumption that the shell model, with its
non-overlapping nucleons, is a valid starting point and therefore should
not be applied to the situation for which the interparticle spacing is
less than the diameter of a nucleon.

We note 
that the extension of this model to the 3-quark
system yields essentially the same results. This is because the
harmonic interaction $r^2=\sum_{i<j}^3 (\vec r_i-\vec r_j)^2=
3(\rho^2+\lambda^2)$ where $\vec\rho=(\vec r_1-\vec r_2)\sqrt2$ and
$\vec\lambda=(\vec r_1+\vec r_2 -2 \vec r_3)/\sqrt6.$
Using the same operator in $H_1$ once again leads to the result that for
harmonic oscillator models the frequency in the nuclear medium is
less than that in free space.

Let us examine the accuracy of our first order (fo)
treatment of Eq.(\ref{deltao}).
In first order $\delta<r^2>_{fo}={-2V_0\rho(0)\over 3\omega_0}<r^2>$
compared with a model exact result of $\delta<r^2>
=\left((1+{4V_0\rho(0)\over 3\omega_0})^{-{1/2}}-1)\right)$.
With our typical parameter, the first order increase in the mean
square radius is 13\%, while the exact result is a shift of 17\%.

The EM form factor in the medium obtained from Eq.(\ref{f21}) is given by
\begin{equation}
F(Q^2)=\left[ 1-2\frac{V_0}{\Delta E }\frac{Q^2}
{6\mu\omega_0}\right] F_0(Q^2),\label{t3}
\end{equation}
where $F_0(Q^2)=\exp(-Q^2/16\mu\omega_0)$ is the free form factor.
Eq.(\ref{t3})
compares favorably with the exact solution with corrections of order
$(V_0/\Delta E )^2$.  Some caution should be exercised, however, in
the application of the approximate solution to large momenta, i.e.,
$Q^2\sim 6\mu\omega_0 \Delta E / V_0$, where the coefficient of the
$(V_0/\Delta E)^2$ correction becomes significant.  It is also
important to note that the interaction with the medium has produced an
increase in the mean square radius, $<r^2>\approx r_0^2[1+\frac{4}{3}
(V_0/\Delta E)]$, as is reflected in both the exact and
the approximate form factor.

\subsection{Coulomb binding}

The harmonic oscillator is not expected to describe situations
involving high momentum transfer.
Indeed we may recall the asymptotic expansion:
\begin{equation}
\lim _{Q^2\to\infty}F_0(Q^2)\approx \left.
{-8\pi\over Q^4} {d \phi_0^2\over dr} \right|_{r=0}
+\left. {16\pi\over Q^6} {d^3\over dr^3}  \phi_0^2\right| _{r=0}.
\label{thm0}
\end{equation}
This power-law dependence gives larger results than the Gaussian form
and holds unless the potential is an analytic function
of $r^2$.  The only relevant example of such is the harmonic oscillator
force of the previous sub-section.
It is useful to obtain the general expression for the
medium-induced change in the form factor $\Delta
F(Q^2)\equiv F(Q^2)-F_0(Q^2)$. To first-order in $H_1$ this is
given by
\begin{equation}
\Delta F(Q^2)=2\sum_{n\ne0}{<\phi_0|e^{i\vec Q\cdot {\vec r\over
2}}|\phi_n><\phi_n|H_1|\phi_0>\over E_0-E_n}.
\label{efo}
\end{equation}
One may then define the state vector $|\chi>$ such that
\begin{equation}
(E_0-H^{rel}_0)|\chi>=(1-|\phi_0><\phi_0|)H_1|\phi_0>,
\label{chi}
\end{equation}
with $<\chi|\phi_0>=0$. Then
\begin{equation}
\Delta F(Q^2)=2 <\phi_0|e^{i\vec Q\cdot {\vec r\over 2}}|\chi>,
\label{delf}
\end{equation}
and
\begin{equation}
\lim_{Q^2\to\infty}\Delta F(Q^2)=\left. {-8\pi\over Q^4} {d\over
dr}(\phi_0\chi)\right| _{r=0}
+\left. {16\pi\over Q^6} {d^3\over dr^3}  (\phi_0\chi)\right| _{r=0}.
\label{thm}
\end{equation}
Our purpose here is to compare the exact
first-order result of Eq.(\ref{chi})
with that of the closure approximation of
Eq.(\ref{f21}).

The above equations are general, but we can gain some
understanding if we specify to the Coulomb Hamiltonian
\begin{equation}
H^{rel}_0=\frac{p^2}{2\mu } -
{e^2/r},
\label{coul}
\end{equation}
with $e^2=4\alpha_s /3$ to simulate the strong color electric
force.
The ground state wave function is given by
\begin{equation}
\phi_0(r)={1\over (\pi a_0^3)^{1\over 2}} e^{-{r\over a_0}},
\end{equation}
with  $a_0=1/(e^2\mu)$.
The form factor for this state is
\begin{equation}
F_0(Q^2)=\left[ 1+{Q^2 a_0^2\over 16}\right] ^{-2}.
\end{equation}

The Hamiltonian of Eq.(\ref{coul}) can be used in Eq.(\ref{chi})
to obtain a solvable differential equation.  The result is
the function $\chi(r)$ given by
\begin{equation}
\chi(r)={V_0\rho(0)\mu a_0^{1/2}\over 3 \sqrt{\pi}} e^{-r/a_0}
[11/2-(r/a_0)^2-(r/a_0)^3/3].
\label{chisol}
\end{equation}
One may use Eqs.(\ref{thm}) and
 (\ref{chisol}) to immediately find that,
at large values  of $Q^2$, $\Delta F(Q^2)$ varies
as $Q^{-4}$ and is negative.
In this model, the high $Q^2$ form factor is dominated by the point
like configuration,
so we may say that the point like configuration is suppressed in the nuclear
medium.
It is not difficult to
use Eqs. (\ref{delf})
and (\ref{chisol}) to obtain an exact expression for $\Delta
F(Q^2)$. However it is more instructive to use the asymptotic
expansion of Eq.(\ref{thm0}) to get
$\Delta F(Q^2)$
as an expansion in powers of $Q^{-2}.$
Then one finds
\begin{equation}
\lim_{Q^2\to\infty}\Delta F(Q^2)={V_0\rho(0)\mu\over 3\;a_0^2\;Q^4}
\left[ 88-{544\over Q^2\;a_0^2}\right] .
\label{high}
\end{equation}
Recall that $V_0$ is a negative  quantity, so that $\Delta F$ is
also negative.
The closure approximation to the change of the form factor,
$\Delta F_{clos}(Q^2)$ can be obtained from Eq.(\ref{f21}) as
\begin{equation}
\Delta F_{clos}(Q^2)={2 V_0\rho(0)\over <r^2> \Delta E}
(-4\bigtriangledown_Q^2)F_0(Q^2),
\end{equation}
which may be evaluated asymptotically as
\begin{equation}
\Delta F_{clos}(Q^2)\approx -{8\;V_0\rho(0)\over 3\Delta E}
\left({16\over Q^2\;a^2_0}\right)^3.
\label{clos}
\end{equation}
It is clear that the closure
approximation can not be valid unless $\Delta E$ is taken to be a
function of $Q^2$. We may equate the $\Delta F's$ of Eqs.(\ref{clos})
and (\ref{high}) to determine the ``correct" value of  $\Delta
E$. The asymptotic result is
\begin{equation}
\Delta E={-4096\over \mu a_0^2} [11Q^2a_0^2-68]^{-1}.
\end{equation}
This means that the magnitude of the $\Delta E $ decreases as
$Q^2$ increases, so that using the closure approximation with a
fixed value of $\Delta E $ underestimates the effects of the
medium modifications. This is the principal result of this
subsection.

\subsection{Harmonic Oscillator plus Coulomb}

A more interesting model is one which
includes
both a
confining and a Coulomb type $1/r$ term in the Hamiltonian.
Thus we consider
\begin{equation}
H^{rel}_0=\frac{q^2}{2\mu }+\frac{1}{2}\mu \omega_0 ^2r^2 -\frac{4}{3}
\frac{\alpha _s}{r}. \label{t4}
\end{equation}
This model was considered in Refs.\cite{FMS92,FMS931} in which it  was
shown that including the Coulomb term leads to
PLC dominance
of the form factor. This occurs even though the Coulomb term
causes only a small effect in the computed energy.

For the calculations performed here the parameters are taken to be
$\alpha _s=0.1$, $\mu = 300\mbox{MeV}/c^2$ and $\hbar \omega_0 =
390$MeV.  Fig.\ref{fig3} shows how the ground state wave function
in free space is influenced by the $1/r$ term in the Hamiltonian. The
attraction enhances the wave function at the origin and changes the shape away
from the Gaussian. This causes a $1/Q^4$ behavior in the  form factor.
Next, Fig. \ref{fig4} compares the full wave function with and without the
effects of the medium. We see that the medium causes a reduction of
the short distance wave function. The closure approximation to the
wave function in the medium is compared with the exact
calculation in Fig. \ref{fig5}.
The numerical
results for the medium-modified
form factor are shown in Fig.\ref{fig6} along with the closure
result and the free
form factor, from which one concludes that closure works well and
in fact slightly underestimates the exact result.  This is consistent
with the conclusions of the previous subsection.

It is also useful to  try to understand these results using analytic
techniques.
This can be done if one works to first order in $\alpha_s$. First
consider the free case. One may find the form factor using
an equation like Eq. (\ref{efo}), but with the operator $H_1$
replaced by $-4\alpha_s/3r$.  The result is
\begin{equation}
F_0(Q^2)=e^{-z_0}\left( 1+ {4\alpha_s/3\over (\sqrt{\pi}/2)}
\sqrt{\mu/\omega_0}f(z_0)\right) .
\label{anal}
\end{equation}
with
\begin{equation}
f(z_0)=\sum_{n=1}^\infty {1\over n} {1\over 2n+1}{1\over
n!}z_0^n,
\end{equation}
where
\begin{equation}
z_0\equiv Q^2/(16\mu\omega_0).
\label{z0}
\end{equation}
The second term of  Eq.(\ref{anal}) leads to a $(\mu\omega_0)^2/Q^4$ behavior
of the form factor which dominates over the exponential term for
values of $z_0$ greater than two or so.
Then the free form factor  varies as
$\omega_0^{3/2}$.
The medium modification is quite easy to implement. Simply
replace $\omega_0$ by $\omega$ of Eq.(\ref{omega})
 in the formulae
(\ref{anal})-(\ref{z0}).  We have seen in Eq.(\ref{omega}) that $\omega= 0.86
\omega_0$ so that the form factor is reduced by a factor of 0.79.
This is a  significant suppression. The results of this perturbative analysis
are shown in Fig.\ref{fig7}.  Comparison with Fig.\ref{fig6} shows that
the perturbative analysis is in good agreement with the full result.

\section{Relativistic Constituent Quark Model of The nucleon}

The influence of medium modifications
on computed matrix elements and form factors
are examined using non-relativistic models in the
previous sections. The need for a relativistic formulation is
clear; we wish to compute form factors at momentum transfers
$Q^2>1$ GeV$^2$, and we wish to be able to distinguish between
electric and magnetic effects.

Here we study the effects of color neutrality on the light front
nucleonic
model wave functions of Schlumpf\cite{fs1,fs2,fs3}.
 The first
formulation
of such a light
front relativistic quark model  was
presented by Terent'ev and Beretskii\cite{bere76,bere77}.
Many authors  
\cite{FS81,azna82,jaus90,chun91,cc,dzie88,webe87,sc,gm,Simula,ss1,keister,kone90,jaco90,hill9
0}
have
contributed to the development of this model.
We use Schlumpf's model because his power-law wave   
functions lead to a reasonably good description of the
proton electromagnetic from factors, $G_E$ and $G_M$, at all of the
$Q^2$ where data are available\cite{fs1,fs2,fs3}.

In a quantum mechanical relativistic theory the commutation
relations between  the ten generators of the Poincar$\acute{e}$
group must be respected. The light front approach is
distinguished by the feature that the maximal number of
seven generators are
of kinematical character (do not contain interactions).
Another feature involves the use of the light front variables
$p^\pm\equiv p^0 \pm p^3$, so that the Einstein mass relation
$p_\mu p^\mu=m^2$ can be expressed as
\begin{equation}
p^-=((p_\perp^2+m^2)/p^+,
\end{equation}
where $p_\perp\equiv(p^1,p^2)$.
Susskind \cite{suss68} noted that this equation is similar to the
non-relativistic kinetic energy if one interprets the variable
$p^+$ to be a relativistic version of the mass. Thus one can use
relative momenta for systems involving several particles, with
the result that
the wave function is  a simple
product of a function involving only relative momenta with a
separate function carrying information about the motion of the
center of mass. One may also employ the Melosh
transformation\cite{melo74} to construct states that are
eigenfunctions of the total angular momentum and its third
component.

The light front dynamics have another important feature,  stressed in
Refs.\cite{BL80,FS81}, that the diagrams with quarks created out of
or annihilated into the vacuum do not contribute. Furthermore,
one need only consider three quark components of the nucleon, if
one is computing the matrix elements of ``good "
operators\cite{good}.

Schlumpf's model is well-documented\cite{fs1,fs2,fs3}.
We reproduce the relevant
features here for the sake of clarity, following his thesis closely.
It is necessary to
express the ten generators of the Poincar\'{e} group $P_\mu$ and
$M_{\mu\nu}$ in terms of dynamical variables to
specify the dynamics of a many-particle system.
The kinematic subgroup
is the set of generators that are independent of the interaction. There
are five ways to choose these subgroups \cite{leut84}.
Usually a physical state is defined
at fixed $x^0$, and the corresponding hypersurface is left
invariant under the kinematic subgroup.

The light-front formalism is specified by the invariant
hypersurface $x^+ = x^0+x^3 =$ constant. The following notation is used: The
four-vector is given by $x = (x^+,x^-,x_\perp)$, where $x^\pm = x^0
\pm x^3$ and $x_\perp=(x^1,x^2)$.
Light-front vectors are denoted by an arrow $\vec x =
(x^+,x_\perp)$, and they are covariant under kinematic Lorentz
transformations \cite{chun88}. The three momenta $\vec p_i$ of the quarks
can be transformed to the total and relative momenta to facilitate
the separation of the center of mass motion
\cite{bakk79} as
\begin{eqnarray}
\vec P&=&\vec p_1+\vec p_2+\vec p_3, \quad \xi={p_1^+\over p_1^++p_2^+}\;,
\quad
\eta={p_1^++p_2^+\over P^+}\;,\nonumber\\
&&\\
q_\perp&=&(1-\xi)p_{1\perp}-\xi p_{2\perp}\;, \quad
K_\perp =(1-\eta)(p_{1\perp}+p_{2\perp})-\eta p_{3\perp}\;.\nonumber
\end{eqnarray}
Note that the four-vectors are not conserved, i.e. $p_1+p_2+p_3\not= P$.
In the light-front dynamics the Hamiltonian takes the form
\begin{equation}
H={P^2_\perp +\hat M^2 \over 2P^+}\;,
\end{equation}
where $\hat M$ is the mass operator with the interaction term $W$
\begin{eqnarray}
\hat M &=&M+W\;, \nonumber\\
M^2&=&{K_\perp^2\over \eta(1-\eta)}+{M_3^2\over \eta}+{m_3^2\over 1-\eta},
\label{eq:2.3} \\
M_3^2&=&{q_\perp^2\over \xi (1-\xi)}+{m_1^2 \over \xi}+{m_2^2\over 1-\xi}\;,
\nonumber
\end{eqnarray}
with $m_i$ being the masses of the constituent quarks. To get a clearer
picture of $M$ we transform to $q_3$ and $K_3$ by
\begin{eqnarray}
\xi&=&{E_1+q_3\over E_1+E_2}\;, \quad \eta={E_{12}+K_3\over E_{12}+E_3}\;,
\nonumber\\
&&\\
E_{1/2}&=&({\bf q}^2+m_{1/2}^2)^{1/2}\;,\quad
E_{3}=({\bf K}^2+m_{3}^2)^{1/2}\;,\quad
E_{12}=({\bf K}^2+M_{3}^2)^{1/2}\;,\nonumber
\end{eqnarray}
where ${\bf q}=(q_1,q_2,q_3)$, and ${\bf K}=(K_1,K_2,K_3)$.
The expression for the mass operator is now simply
\begin{equation}
M=E_{12}+E_3\;, \quad M_3=E_1+E_2\;.
\end{equation}

The use of
light front variables enables one to
separate the center of mass motion from the
internal motion. The internal wave function $\Psi$ is
therefore a function of the
relative momenta ${\bf q}$ and ${\bf K}$. The function $\Psi$ is a product
$\Psi=
\Phi\chi\phi$, with $\Phi=$ flavor, $\chi=$ spin, and $\phi=$
momentum distribution. The
color wave function is antisymmetric.

The angular momentum ${\bf j}$ can be expressed as a sum of orbital and
spin contributions
\begin{eqnarray}
{\bf j}=i\nabla_{\bf p}\times {\bf p}+\sum_{j=1}^3 {\cal R}_{Mj}{\bf s}_j \;,
\end{eqnarray}
where ${\cal R}_M$ is a Melosh rotation \index{Melosh rotation}
acting on the quark spins
${\bf s}_j$, which has the matrix representation (for two particles)
\begin{equation}
\left< \lambda' |{\cal R}_M(\xi,q_\perp,m,M)|\lambda\right> =
\left[ {m+\xi M-i{\bf \sigma}\cdot({\bf n}\times {\bf q})\over
\sqrt{(m+\xi M)^2+q_\perp^2}}\right]_{\lambda'\lambda}
\end{equation}
with ${\bf n}=(0,0,1)$.
The effects of the Melosh rotation are to significantly increase the computed
charge radius\cite{cc}.

The operator ${\bf j}$ commutes with the mass operator $\hat M$; this is
necessary and sufficient for Poincar\'e-invariance of the bound state.
In particular, $j^2|\Psi,\uparrow>=3/4|\Psi,\uparrow>$ and
$j_z|\Psi,\uparrow>=1/2|\Psi,\uparrow>$.
The angular momentum operator is in terms of relative coordinates given by
\begin{eqnarray}
{\bf j}&=& i\nabla_{\bf K}\times {\bf K}+{\cal R}_M(\eta,K_\perp,M_3,M)
{\bf j}_{12}+{\cal R}_M(1-\eta,-K_\perp,m_3,M){\bf s}_3\;,\nonumber\\
&&\\
{\bf j}_{12}&=& i\nabla_{\bf q}\times {\bf q}+{\cal R}_M(\xi,q_\perp,m_1,M_3)
{\bf s}_1+{\cal R}_M(1-\xi,-q_\perp,m_2,M_3){\bf s}_2\;.\nonumber
\end{eqnarray}
The orbital contribution
does not contribute for the ground state baryon octet, so
that
\begin{eqnarray}
{\bf j}&=&\sum {\cal R}_i{\bf s}_i\;,\nonumber\\
{\cal R}_1&=&{1\over \sqrt{a^2+K_\perp^2}\sqrt{c^2+q_\perp^2}}
\pmatrix{ac-q_RK_L&-aq_L-cK_L\cr
         cK_R+aq_R&ac-q_LK_R}\;,\nonumber\\
{\cal R}_2&=&{1\over \sqrt{a^2+K_\perp^2}\sqrt{d^2+q_\perp^2}}
\pmatrix{ad+q_RK_L&aq_L-dK_L\cr
         dK_R-aq_R&ad+q_LK_R}\;,\label{eq:melosh}\\
{\cal R}_3&=&{1\over \sqrt{b^2+K_\perp^2}}\pmatrix{b&K_L\cr
         -K_R&b}\;,\nonumber
\end{eqnarray}
with
\begin{eqnarray}
a&=&M_3+\eta M\;,\quad b=m_3+(1-\eta)M\;,\nonumber\\
c&=&m_1+\xi M_3\;, \quad d=m_2+(1-\xi)M_3\;,\nonumber\\
q_R&=&q_1+iq_2\;,\quad q_L=q_1-iq_2\;,\\
K_R&=&K_1+iK_2\;,\quad K_L=K_1-iK_2\;.\nonumber
\end{eqnarray}
The momentum wave function
can be chosen as a function of $M$ to fulfill the requirements of
spherical and
permutation symmetry.
The $S$-state orbital function $\phi(M)$ is approximated by either
\begin{equation}
\phi(M)=N\exp\left[-{M^2\over 2\beta_G^2}\right] \qquad\hbox{or}\qquad
\phi(M)={N'\over (M^2+\beta^2)^{3.5}}\;,
\label{eq:2.34}
\end{equation}
which depend on two free parameters, the constituent quark mass and the
confinement scale parameter $\beta$. The first
function is the conventional choice used in spectroscopy,
but it has a too strong falloff for
large values of the four-momentum transfer.
We use Schlumpf's
 parameters $\beta_G$=0.56 GeV, $\beta$=0.607 GeV, and the constituent quark
mass, $m_i$=0.267 GeV.

The total wave function for the proton is given by
\begin{eqnarray}
p&=&{-1\over\sqrt 3}\left(uud\chi^{\lambda3}+udu\chi^{\lambda2}
+duu\chi^{\lambda1}\right)\phi\;,\nonumber\\
\end{eqnarray}
with
\begin{eqnarray}
\chi^{\lambda3}_\uparrow&=&{1\over\sqrt 6}(\downarrow\uparrow\uparrow+
\uparrow\downarrow\uparrow-2\uparrow\uparrow\downarrow),\nonumber\\
\chi^{\lambda3}_\downarrow&=&{1\over\sqrt 6}(2\downarrow\downarrow\uparrow-
\downarrow\uparrow\downarrow-\uparrow\downarrow\downarrow)\;.
\label{eq:spinfunction}
\end{eqnarray}
The spin wave functions $\chi^{\lambda2}$ and $\chi^{\lambda1}$ are the
appropriate permutations of $\chi^{\lambda3}$.
The spin-wave function of the $i$th quark is given by
\begin{equation}
\uparrow={\cal R}_i\pmatrix{1\cr 0} \hbox{  and  }
\downarrow={\cal R}_i\pmatrix{0\cr 1}\;.
\end{equation}

We now turn to the calculation of the proton form factors.
The electromagnetic current matrix element can be written
in terms of two form factors taking into account current and parity
conservation:

\begin{equation}
\left< N,\lambda ' p' \left| J^\mu \right| N,\lambda p\right> =
\bar u_{\lambda '}(p') \left[ F_1(Q^2)\gamma^\mu + {F_2(Q^2) \over
2 M_N}i\sigma^{\mu\nu}(p'-p)_\nu \right] u_\lambda (p)
\label{eq:3.1}
\end{equation}
with momentum transfer $Q^2=-(p'-p)^2$ and $J^\mu = \bar q \gamma^\mu q$.
For $Q^2 = 0$ the form factors $F_1$ and $F_2$ are respectively equal
to the charge and the anomalous magnetic moment in units $e$ and
$e/M_N$, and the magnetic moment is $\mu = F_1(0) + F_2(0)$.
The Sachs form factors are defined as
\begin{equation}
G_E = F_1 - {Q^2 \over 4M_N^2}F_2\;, \quad\hbox{and}\quad
G_M = F_1 + F_2\;,
\end{equation}
and the charge radii \index{charge radius} of the nucleons are
\begin{equation}
\left< r_i^2 \right> = -6{dF_i(Q^2) \over dQ^2}\Bigg\vert_{Q^2=0}\;,
\quad\hbox{and}\quad
\left< r_{E/M}^2 \right> = -{6 \over G_{E/M}(0)}{dG_{E/M}(Q^2) \over dQ^2}
\Bigg\vert_{Q^2=0}\;.
\end{equation}
The form factors can be expressed in terms of the $+$ component of
the current:
\begin{eqnarray}
F_1(Q^2) &=&{1 \over 2P^+}\left< N,\uparrow\left| J^+\right| N,
\uparrow\right>\;,\nonumber\\
&&\\
Q_\perp F_2(Q^2) &=&-{2M_N \over 2P^+}\left< N,\uparrow\left|
J^+\right| N,\downarrow\right>\;.\nonumber
\end{eqnarray}

The form factors are calculated from the diagrams of
Fig.\ref{fig8}, using
the ``good" current $J^+$
so that no terms with $q\bar q$ pairs are involved. Schlumpf's
result is
\begin{eqnarray}
F_1(Q^2) &=& {N_c \over (2\pi)^6} \int\!d^3\!q d^3K
\left( {E_3'E_{12}'M \over E_3E_{12}M'} \right)^{1/2}\phi^\dagger(M')\phi(M)
\nonumber\\
&&\times\sum_{i=1}^3 F_{1i}\left<\chi_\uparrow^{\lambda i}|\chi_
\uparrow^{\lambda i}\right> \nonumber\\
&&\label{eq:bracket1}\\
Q_\perp F_2(Q^2) &=& -2M_N{N_c \over (2\pi)^6} \int\!d^3\!q d^3K
\left( {E_3'E_{12}'M \over E_3E_{12}M'} \right)^{1/2}\phi^\dagger(M')\phi(M)
\nonumber\\
&&\times\sum_{i=1}^3 F_{1i}\left<\chi_\uparrow^{\lambda i}|\chi_
\downarrow^{\lambda i}\right>
\nonumber
\end{eqnarray}
with $i=(uud)$ for the proton.  Here the prime indicates the absorption of
the momentum transfer as $K'_\perp =K_\perp +\eta Q_\perp$ and $q'_\perp =
q_\perp $.
The factors $F_{1u}$ and $F_{1d}$ are the charges
 of the
$u$ and $d$ quarks.
We also consider $G_A(Q^2)$.  We take the hadronic axial-vector current
to be
\begin{equation}
A_\mu = \bar u \gamma_\mu \gamma_5 d \; .
\end{equation}
The form factor of interest is given by
\begin{eqnarray}
2P^+ G_A(Q^2) &=&\left< B',\uparrow \left| A^+\right|
B,\uparrow\right>\;.\nonumber\\
\label{eq:formfactors}
\end{eqnarray}

The final step is to specify the operator $r^2$ for this relativistic
model. We use
\begin{equation}
r^2=3(\rho^2+\lambda^2)
\end{equation}
where $\vec \rho$ and $\vec \lambda$ are canonically conjugate to
the momenta $\vec q$ and $\vec K$:
\begin{equation}
\vec \rho={\vec r_1-\vec r_2\over \sqrt {2}} \; \; \; \; \;
\vec \lambda={2\over \sqrt 6}\left(\xi\vec r_1+(1-\xi)
\vec r_2-\vec r_3\right).
\end{equation}
Note that $\vec\rho$ and $\vec\lambda$ reduce to the usual
three-body variables in the non-relativistic limit of
$\xi\to {1\over  2}$.

\section{Results}

It is worth while to begin by discussing whether or not
point-like configurations occur in the
relativistic models we employ.  We do this by defining a quantity
$r^2(Q^2)$ as
\begin{eqnarray}
r^2(Q^2)&\equiv &\frac{\left< N,\uparrow\left| r^2 J^+\right| N,
\uparrow\right>}{\left< N,\uparrow\left| J^+\right| N,
\uparrow\right>}\nonumber \\
&=&{N_c \over (2\pi)^6 F_1(Q^2)}\int\!d^3\!q d^3K
\left( {E_3'E_{12}'M \over E_3E_{12}M'}
\right)^{1/2}
\sum_{i=1}^3 F_{1i}\left<\phi(M'),\chi_\uparrow^{\lambda i}|r^2|\chi_
\uparrow^{\lambda i},\phi(M)\right> .\label{r^2}
\end{eqnarray}
This quantity should not be confused with the transverse size, $b^2(Q^2)$,
used in studies of color transparency.
The production of a
point-like configuration at large momentum transfer by the electromagnetic
current is signaled by the vanishing of the transverse size
for large $Q^2$.  Nevertheless, since we are interested here in the
properties of a nucleon which is not necessarily moving at a large
momentum with respect to the medium, it is $r^2(Q^2)$, as
defined in (\ref{r^2}), which enters our calculations.  For comparison,
we may define the quantity $b^2(Q^2)$ from (\ref{r^2}) by including only
a single transverse component of the operator $r^2$.
We see from Fig. \ref{fig8} that both the power-law and the
Gaussian wave functions display significant reductions
of the quantity $r^2(Q^2)$ for increasing $Q^2$, however only the
power-law wave function is suppressed in the case of the transverse size,
$b^2(Q^2)$.  This behavior has been noted previously\cite{FMS92}, and
exemplifies the feature that power-law wave functions have
point-like configurations in the transverse variables, but that such are absent
in Gaussian  wave functions. 
We shall use only the power law form, unless otherwise noted, because of
its ability to reproduce data.

Comparisons between the free form factors and the medium
modified ones are  given in Figs. \ref{fig10},\ref{fig11}. The quantity
$G_E$ is suppressed at low momentum transfer, but $G_M$ is not.
This is due to the relativistic nature of our model. Indeed, the
magnetic moment is increased by about 5\%. This small change is not enough to
cause disagreement with existing nuclear phenomenology.
At higher momentum transfers both form factors are inhibited by
about 10\%. This is a significant effect, but not large enough to
disagree with Jourdan's analysis\cite{jo95}.

This is more clearly seen by computing the
nuclear response functions
for the inclusive $(e,e')$\ cross section.
(See for example \cite{kim95}.)  The excitation energy
is
$\omega$\ and the three
momentum transfer is ${\bf q}$ so that
\begin{eqnarray}
{d^2\sigma\over d\Omega dE}=
\sigma_M\Bigl[{Q^4\over {\bf q}^4} R_L({\bf
q},\omega)+\Bigl({Q^2\over
2{\bf q}^2} +
{\rm tan}^2{\theta\over 2}\Bigr) R_T({\bf q},\omega)\Bigr]\
,\label{cross}
\end{eqnarray}
with the Mott
cross-section $\sigma_M=\alpha^2{\rm cos}^2(\theta/2)/4E^2{\rm
sin}^2(\theta/2)$.  Here $Q^2=- q_\mu^2={\bf q}^2-\omega^2$\
and $\theta$\ is the
scattering angle.  The longitudinal $R_L$\ and transverse $R_T$\
response
functions are calculated in the relativistic Fermi gas approximation at
a density $\rho=2 k_F^3/3\pi^2$,
\begin{eqnarray}
R_L&=&-{2\over \pi\rho} {\rm Im}
(Z\Pi_{00}^p+N\Pi_{00}^n)\label{long}\ ,\\
R_T&=&-{4\over \pi\rho} {\rm Im}
(Z\Pi_{22}^p+N\Pi_{22}^n)\ ,
\end{eqnarray}
for a target with $Z$\ protons and $N$\
neutrons.  (Note, {\bf q} is assumed to be along the ${\hat {\bf
1}}$
axis so the subscript
22 refers to a transverse direction.)  Here the Fermi momentum is taken to
be $k_F=260$ MeV, which is appropriate for $^{56}$Fe.

The polarization $\Pi$\,
\begin{equation}
\Pi_{\mu\nu}^i(q,\omega)= -i \int
{d^4p\over (2\pi)^4} {\rm Tr}[ G(p+q)\Gamma_\mu^i G(p)
\Gamma_\nu^i ]\ ,
\label{pol}
\end{equation}
is calculated with the nucleon Greens
function $G(p)$
\begin{equation}
G(p) = {p_\mu \gamma^\mu + M_{\bf p} \over 2 E_{\bf p}}\
\Biggr[{\Theta (k_F
-|{\bf p}|) \over p_0-E_{\bf p}-i \epsilon}
+{\Theta (|{\bf p}|-k_F)
\over p_0-E_{\bf p}+i\epsilon}\Biggl]\ ,
\end{equation}
and the electromagnetic vertex $\Gamma$,
for $i=$\ p (proton) or n (neutron).  We assume the
electromagnetic vertex has the
form,
\begin{equation}
\Gamma_\mu^i = F_1^i \gamma_\mu + F_2^i{i\sigma_{\mu\nu}q^\nu
\over 2M}\label{ver}
\end{equation}
even for off-shell nucleons in the medium. We show results
comparing using $F_i$ computed in free space and in the medium in
Figs. \ref{fig12} and \ref{fig13}.

It is immediately evident from Fig.\ref{fig12} that modifications to
the magnetic form factor $G_M$ have no effect on the longitudinal response,
while the modifications to the electric form factor $G_E$ lead to
a suppression.  This effect is due to the larger charge radius in the medium.
The situation is quite different in the case of the transverse response shown
in Fig.\ref{fig13}.  There it is seen that modifications to the electric
form factor have no effect, while modifications to the magnetic form factor
lead to only a minor suppression.  In this case the increase in the
magnetic moment, as shown in Fig.\ref{fig11}, tends to cancel the effect of the
increased radius.

We also study the effects of using the form factors of Figs. \ref{fig10} and
\ref{fig11} in computing the $(e,e'p)$ cross sections for finite nuclei.
In this case, absorption effects emphasize the role of the nucleon surface and
the influence of color neutrality is about 60\% smaller than shown in
Figs. \ref{fig12} and \ref{fig13}.

Our results for the medium modifications on the form factor $G_A$
are shown in Fig.\ref{fig14}. Once again there is about a
10\% reduction. This effect could be observed in
neutrino-nucleus scattering, in the ($p,n$) reaction, or
parity violating electron scattering\cite{barbaro94}.

We now turn to  computing the valence structure functions in the
medium and in free space. One advantage of the light front
formalism is that the
wave function is closely related to the valence structure function
$F_2^{val}$:
\begin{equation}
F_2^{val}(x_{Bj})=\int d^3q d^3K \Psi^{\dagger }\Psi
\delta\left( x_{Bj}-{p_3^+\over P^+}\right).
\end{equation}
Thus the operator ${\cal O}$ of Eq.~(\ref{f21}) is the delta
function which sets the plus-momentum of the quark equal to
$x_{Bj} P^+.$
This is meant to be taken at some momentum scale $\sim 1$ GeV. We may
compute this quantity for the free and medium modified wave
function. There are a host of other effects of the medium
including Fermi motion, shadowing, pions in the medium, nuclear
correlations, six quark bags, etc. (see the reviews
\cite{EMCrefs}.)  Our concern here is  to assess the effects of color
neutrality.  So we do not include these effects and do not compare
 our results with data.

The free version is shown in Fig.\ref{fig15}, where the expected shape
is obtained. The results for the ratio of medium modified to
free structure functions
using both the power law and Gaussian forms are shown
in Fig.\ref{fig16}. This
shows that suppression of point like configurations does indeed
lead to suppression of the valence structure function at large
values of $x_{Bj}$.
 The normalization of the wave function
ensures that the integral $\int dx_{Bj} F_2(x_{Bj})$ is unity
whether the free or modified wave function is used.  Thus one
expects regions in which the ratio is bigger and smaller than
unity. However,
the result that there is suppression at large $x_{Bj}$
is not a trivial consequence of
kinematics and normalization as is seen by comparing with the Gaussian
form.  The Gaussian wave function does not display as much suppression
at large $x_{Bj}$, which indicates the relative importance of large
momentum components, or
point-like configurations, in the power law versus Gaussian
wave functions.
The suppression occuring at large $x_{Bj}$
is consistent with
the relevant features of the data and therefore provides evidence
for the existence of point like configurations.

\section{Summary and Discussion}

We have used the ideas of color neutrality to motivate a
functional form of the central shell model potential:
\begin{equation}
V(r,R)=V_0\rho(R){r^2\over<r^2> },
\end{equation}
where $<r^2>\equiv<\phi_0|r^2|\phi_0>$
 is the nucleonic expectation value of the operator
$r^2$, $V_0\approx-50 $MeV and $\rho(R)$ is the nuclear density
normalized so that $\rho(R=0)=1$.
The
concern here is  with fluctuations, so that we
rewrite the central shell model potential as
\begin{equation}
V(r,R)=V_0\rho(R) +
V_0\rho(R)\quad{r^2-<r^2>\over<r^2> },
\end{equation}
and treat the second term as a perturbation. The effects of this perturbation
can be evaluated,  using Eq.(\ref{fnew2}) (or using the closure approximation
of
Eq.(\ref{f21})) for any nuclear process.  

Solving a set of toy models indicates  that a perturbative  
treatment is valid and furthermore that a closure
approximation may be used to  avoid computing the
complete spectrum of baryonic wavefunctions usually necessary in
perturbation theory. The toy-model results are consistent with 
the notion that
point like configurations  can be suppressed in the nuclear medium.

Light front quantum mechanics, along with a specific model of the
nucleon wave function
\cite{fs1,fs2,fs3}, is next employed to compute
the influence of medium effects at relatively high momentum
transfer. We find that at low values of $Q^2< 1 $ GeV$^2$
  the electric
form factor is suppressed and displays an increased charge radius,
but that while the magnetic radius is also increased so is the magnetic
moment.  This leads to the result that the $(e,e')$ transverse response,
shown in Fig.\ref{fig13}, is largely unaffected by the medium in this context.
This behavior of the transverse response has previously been interpreted
as signaling no change in the magnetic radius --
 contrary to the result obtained
here.  At higher values of $Q^2$ both form factors are
suppressed in the medium as
is $G_A$. These results are  in accord with ideas about the suppression of
the longitudinal response, but are not inconsistent
with the analysis of Jourdan\cite{jo95}.

The results for the medium modifications of the $F_2$ structure
function show a suppression at large values of $x_{Bj}$ for both the
power-law and Gaussian wave functions.  However, the suppression is greater
in the case of the power law, indicating the more pronounced role of
high momentum components or
point-like configurations there.
This is consistent with the works of \cite{FS85} and \cite{FMS92,FMS931} and
more importantly, provides evidence for the existence of point like
configurations.


\acknowledgements

This work was supported by the
Department of Energy under Grants No. DE-FG06-90ER40561
and DE-FG06-88ER40427.  We wish to acknowledge helpful conversations with
Brad Keister.


\begin{figure}
\caption{Possible configurations of  the proton wave function.}
\label{fig1}
\end{figure}

\begin{figure}
\caption{The matrix element of Eq. (12) ($j_{\mu }$ commutes with
$h^{rel}_1$.) }
\label{fig2}
\end{figure}

\begin{figure}
\caption{The influence of the attractive Coulomb potential in Eq.(\ref{t4})
on the wave function is illustrated.}
\label{fig3}
\end{figure}

\begin{figure}
\caption{The influence of the medium on the wave function is illustrated
as described in the text.}
\label{fig4}
\end{figure}

\begin{figure}
\caption{The medium-modified exact wave function
and that calculated in the closure approximation are plotted as a
function of the relative coordinate, $r$.}
\label{fig5}
\end{figure}

\begin{figure}
\caption{The exact medium-modified, free, and the
closure-approximated form factors are plotted as a function of the square
of the momentum transfer.}
\label{fig6}
\end{figure}

\begin{figure}
\caption{Medium modified and free form factors calculated from the analytic
expression given in Eq.(\ref{anal}) are plotted as a function of
the square of the momentum transfer. }
\label{fig7}
\end{figure}

\begin{figure}
\caption{The absorption of momentum by the valence quarks is illustrated. }
\label{fig8}
\end{figure}

\begin{figure}
\caption{The quantities $r^2(Q^2)$ and $b^2(Q^2)$
for  power-law and Gaussian wave functions.}
\label{fig9}
\end{figure}

\begin{figure}
\caption{The free and medium modified electric form factors vs.
the square of the four-momentum transfer, $Q^2$.}
\label{fig10}
\end{figure}

\begin{figure}
\caption{The free and medium modified magnetic form factors vs. $Q^2$.}
\label{fig11}
\end{figure}

\begin{figure}
\caption{The longitudinal response  vs
the energy transfer, $\omega$.}
\label{fig12}
\end{figure}

\begin{figure}
\caption{The transverse response vs. $\omega$.}
\label{fig13}
\end{figure}

\begin{figure}
\caption{The free and medium-modified axial vector form factors vs $Q^2$.}
\label{fig14}
\end{figure}

\begin{figure}
\caption{Proton structure functions are plotted as a function of the
scaling variable $x_{Bj}$ for power-law and Gaussian wave functions.}
\label{fig15}
\end{figure}

\begin{figure}
\caption{The ratio of the medium-modified to free structure function is
plotted as a function of the scaling variable for the power-law and
Gaussian wave functions.}
\label{fig16}
\end{figure}

\end{document}